\documentstyle[11pt,newpasp,twoside,epsf,graphicx]{article}
\def\edcomment#1{\iffalse\marginpar{\raggedright\sl#1\/}\else\relax\fi}
\reversemarginpar

\begin{document}
\title{The Most Massive Clusters in the Universe}
 \author{Oliver Czoske, Jean-Paul Kneib \& S\'ebastien Bardeau}
\affil{Observatoire Midi-Pyr\'en\'ees, 14 avenue \'Edouard Belin, 31400
 Toulouse, France}

\begin{abstract}
  Clusters of galaxies need to be investigated using complementary
  approaches combining all currently available observational
  techniques (X-ray, gravitational lensing, dynamics, SZ) on
  homogeneous samples if one wants to understand their evolution and
  physical properties.  This is particularly important in order to
  relate the observable quantities to the cosmologically important
  cluster mass. We present here a number of on-going projects that aim
  at studying cluster physics for samples based on currently available
  all-sky X-ray surveys such as XBACs, BCS and MACS.
\end{abstract}

\section{Introduction}

Clusters of galaxies are important probes for cosmology. In
hierarchical scenarios for structure formation in the Universe they
are the latest class of gravitationally bound objects to have formed
and their formation and evolution can be observed at fairly low
redshifts, $z \la 1$. The mass function of clusters of galaxies
depends on the mean mass density in the Universe, $\Omega_{\rm M}$, as
well as on the normalisation of the density fluctuation power
spectrum, $\sigma_8$. Measurements of the local cluster mass function
constrain the combination $\Omega_{\rm M}^{0.5}\sigma_8$; the degeneracy
is broken through observations of the evolution of the mass function
with redshift $z$ (e.~g.\ Eke et al.\ 1996). The evolution of the
mass function is strongest, and hence in principle most easily
observable, at the high-mass end.

The mass distribution within clusters depends on the type and
properties of the elusive dark matter. In cold dark matter scenarios,
numerical simulations indicate the existence of a universal dark
matter profile (Navarro, Frenk, \& White 1997), which falls off as
$r^{-3}$ at large radius and has a central cusp of limiting slope
between $-1$ and $-1.5$ (Navarro et al.\ 1997, Moore et al.\ 1998,
Ghigna et al.\ 2000). Warm or self-interacting dark matter would
result in more extended central mass distributions, i.~e.\ a central
flat core (e.~g.\ Spergel \& Steinhardt 2000). Probing the central
mass distribution in clusters on scales of $\sim
10\,h^{-1}\,\mathrm{kpc}$ can thus provide valuable information on the
properties of the dark matter, although flat cores can be mimicked or
created by projection effects or line-of-sight cluster mergers (Czoske
et al.\ 2002).  The slope of the total mass profile is also strongly
affected by the presence of baryons and particularly galaxies (stars)
that suffer collisions unlike CDM particles, so we do not expect that
the actual slope of the mass profile in the very centre follows an
NFW-like profile. This effect is difficult to model and few numerical
simulations have attacked this problem so far. Lensing measurements of
central mass profiles can provide valuable clues on the behaviour of
baryons in cluster size dark matter halos.  Massive clusters are
particularly interesting in this context because their high central
mass density enables them to multiply image suitably placed background
galaxies (strong lensing, see below), thus providing a means to
accurately reconstruct the central mass distribution at high
resolution.

Traditionally, clusters of galaxies have been found through optical
(Abell 1958, Zwicky et al.\ 1961, Las Campanas Distant Cluster Survey,
EDISCs, Red-Sequence Cluster Survey) or X-ray methods. While optical
methods yield larger samples, their selection criteria are not as
closely related to cluster mass as is X-ray luminosity. Furthermore,
optical searches are prone to projection effects. X-ray surveys are
much less affected by projection because the surface brightness varies
as the square of the gas density and thus directly probes gas trapped
in deep potential wells. The most massive clusters are therefore most
reliably selected from X-ray surveys (Fig.\ 1).  Since massive
clusters are rare objects, X-ray surveys need to cover large areas on
the sky, so that (nearly) all-sky surveys such as Reflex (B\"ohringer
et al.\ 2001) or BCS (Ebeling et al.\ 1998) are needed to construct
large samples of very massive clusters, with only the on-going MACS
survey (Ebeling et al.\ 2001) providing sufficient depth to find
massive clusters at high redshift ($z>0.3$). For the future, weak
lensing and Sunyaev-Zeldovich (SZ) surveys will provide new cluster
catalogues.  Whereas weak lensing surveys will select clusters
directly by (projected) mass (Miyazaki et al.\ 2002), their survey
areas will be too small to yield complete samples of the most massive
clusters.  Furthermore, since weak lensing analyses require large
numbers of background galaxies, the clusters found by this method will
probably be restricted to a redshift range of $0.1<z<0.8$ (Ellis
2001). The SZ effect is independent of redshift and thus holds great 
promise for constructing cluster catalogues particularly at high
redshift. However, a large cluster catalogue won't be available until
the all-sky SZ survey conducted by the \textsc{Planck} mission.

Every cluster selection method also provides a way to estimate cluster
masses. The most direct route to the total mass distribution in
clusters of galaxies is provided by the gravitational lens effect
which is sensitive to the total mass independent of the nature of the
matter (dark or baryonic) or its dynamical state. Lensing is however
sensitive to the weighted sum of all mass between the observer and the
source, hence the interpretation of the measured mass as a {\em cluster}
mass can be problematic in the presence of several mass concentrations
along the line of sight (e.~g.\ Czoske et al.\ 2002). Also,
substructure in the vicinity of the cluster can bias masses determined
from gravitational lensing (Metzler et al.\ 2001).

The X-ray emission from clusters of galaxies comes from hot gas
confined within the cluster potential well and is less sensitive to
projection effects than optical and lensing methods. However, in order
to estimate the total cluster mass from X-ray observations
(temperature, surface brightness distribution), one generally assumes
that the gas is in hydrostatic equilibrium and that the cluster is
spherically symmetric. The assumption of hydrostatic equilibrium can
be strongly violated in particular in merging clusters in the presence
of shock waves. Another problem with the interpretation of X-ray
observations is that the observed relation between X-ray luminosity
and X-ray temperature, $L_{\rm X}\propto T_{\rm X}^3$ differs from the
theoretically expected relation $L_{\rm X} \propto T_{\rm X}^2$ if the
gas is heated purely through gravitational collapse, so that other,
non-gravitational heating mechanisms have to be invoked (e.~g.\ Henry,
this volume). In the absence of a complete theoretical understanding
of these heating mechanisms the relations between X-ray observables
and cluster mass have to be calibrated observationally (Smith et al.\
2003). 

Gravitational lensing gives rise to a diverse phenomenology that
allows to probe the mass distribution in clusters over a wide range of
length scales (see the contribution by M. Bartelmann, this volume).
The central parts of clusters are characterised by their large surface
mass density capable of producing multiple images of suitably placed
background galaxies. Positions and flux ratios of these multiple
images allow a detailed reconstruction of the central projected mass
distribution at resolutions of typically $\sim
10\,h^{-1}\,\mathrm{kpc}$. If multiple image systems at different
source redshifts are present in the same cluster, it is possible to
constrain the geometry of the Universe and measure the cosmological
parameters $\Omega_{\rm M}$ and $\Omega_\Lambda$ (Golse, Kneib \&
Soucail 2002). Further away from the cluster centre, the
gravitational shear, i.~e.\ the distortion of background galaxy shapes
introduced by the lens potential, is weak, at best of the order of the
intrinsic scatter in galaxy ellipticities, and hence only measurable
statistically, by averaging over large numbers of background galaxies.
Consequently the spatial resolution of weak shear mass reconstructions
is poor; on the other hand, coherent shear signals can be detected out
to distances of $1\dots2\,h^{-1}\,\mathrm{Mpc}$ from the cluster
centre.

In this contribution we present a number of projects aiming at
studying the properties of the most massive clusters through
gravitational lensing and other methods.

\begin{figure}[htbp]
  \centering
  \resizebox{!}{0.35\textheight}{\plotone{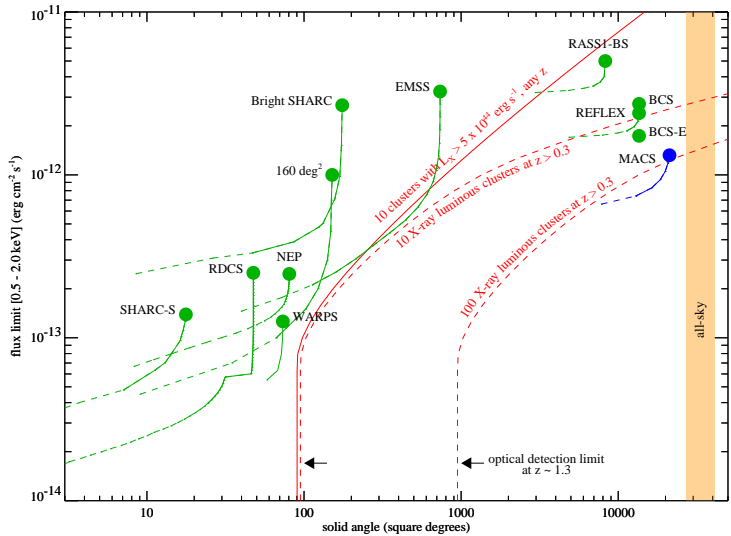}}
  \caption{Synopsis of X-ray surveys for clusters of galaxies
  (courtesy of Harald Ebeling). All-sky surveys like Reflex or BCS are
  needed to find the most luminous and hence most massive clusters of
  galaxies, but only MACS provides the combination of depth and survey
  area to find statistically useful numbers of massive clusters at
  high redshift.}
  \label{fig:X-surveys}
\end{figure}

\section{Panchromatic High-$L_{\rm X}$ $z\sim 0.2$ Survey}

This project aims at compiling as complete data sets as possible for a
homogeneously selected sample of X-ray luminous, hence presumably
massive, clusters of galaxies. In order to minimise evolutionary
effects, the clusters were chosen to lie within a narrow range of
redshifts around $z\approx0.2$. The observational corner stones of the
project include high-resolution imaging with HST/WFPC2, multi-colour
wide-field imaging with the CFH12k camera on CFHT, and X-ray imaging
and spectroscopy with XMM/Newton (Marty et al.\ 2002). The data are
complemented by optical multi-object spectroscopy, as well as near-IR
imaging with UKIRT and VLT.

\subsection{Sample Selection}

The clusters were selected from the XBACs catalogue of Ebeling et al.\ 
(1996). This catalogue is a flux-limited compilation of Abell clusters
identified in the \textsc{Rosat} All-Sky Survey data. While this
catalogue is based on the optically selected Abell catalogue which is
known to be incomplete at high redshift, comparison with the purely
X-ray selected BCS (Ebeling et al.\ 1998) shows that more than 80\% of
the BCS clusters are indeed Abell clusters and included in XBACs.
Sample completeness is a minor concern for this project which rather
aims at compiling a homogeneous data set for a representative cluster
sample. Fig.\ 2 shows the location of this sample
within the XBACs catalogue.

\begin{figure}[htbp]
  \label{fig:sample}
  \centering
  \resizebox{!}{0.3\textheight}{\plotone{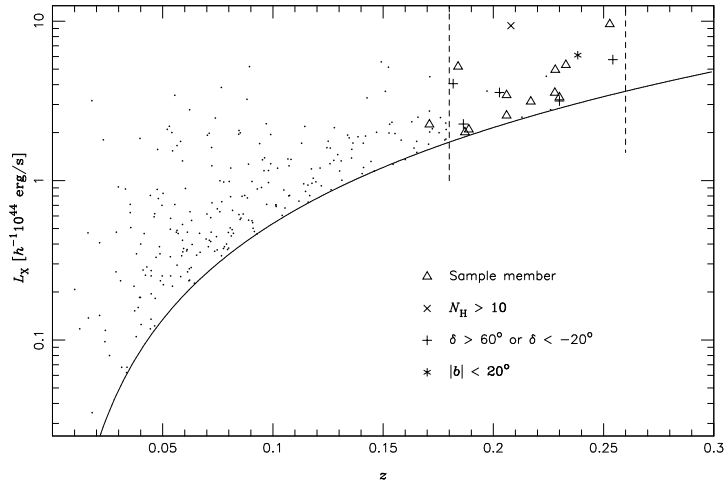}}
  \caption{Location of the $z=0.2$ high-$L_{\rm X}$ sample within
    XBACs. Triangles mark sample members, other symbols mark clusters
    that were excluded due to high hydrogen column density,
    inaccessibility from CFHT (Mauna Kea) or low galactic latitude.
    Abell 2218 was observed in the same manner as the other sample
    members despite being outside the redshift limits $0.18<z<0.26$.
    The solid line marks the flux limit of the XBACs catalogue.}
\end{figure}

\subsection{Strong Lensing}

The excellent resolution of the WFPC2 images of the central parts of
the clusters in our sample allows a detailed study of the giant arc systems
which are present in the majority of the clusters. Eight clusters of
the sample (A~68, A~209, A~267, A~383, A~773, A~963, A~1763,
A~1835) were observed in Cycle 8 (P.~I.\ J.-P. Kneib) through the
F702W filter with three orbits per cluster. Images for the remaining
four clusters were taken from the HST archive.

A strong lensing model for Abell 383 was presented by Smith et al.\
(2001). This regular cluster shows a multitude of giant arcs and
arclets which allowed a detailed determination of the radial density
profile of the cluster and to assess the influence of individual
galaxies which break up the arcs to the south of the cD
galaxy. The presence of two radial arcs at distances of $1\farcs5$
and  $5\arcsec$ (corresponding to $3\,h^{-1}\,\textrm{kpc}$
and $10\,h^{-1}\,\mathrm{kpc}$ at
the cluster redshift) imposes strong constraints on the slope of the
density profile at these distances. Fig.\ 3 shows
the reconstructed three-dimensional density profile in Abell 383;
approaching the centre the profile first flattens as expected for a
CDM type profile, reaching a slope of $-1.3\pm0.04$ at the position of
the outer radial arc, but then steepens again, reaching a slope of
$-1.5\pm0.04$ at the position of the inner radial arc. Similar profiles
are observed in most of the other clusters in our sample (Smith et
al.\ 2003). The steepening is most likely due to the
contribution of the stars in the cD galaxies, which is not accounted
for in numerical simulations involving only dark matter. 

\begin{figure}
  \hspace*{0.05\textwidth}
  \resizebox{0.3\textwidth}{!}{\plotone{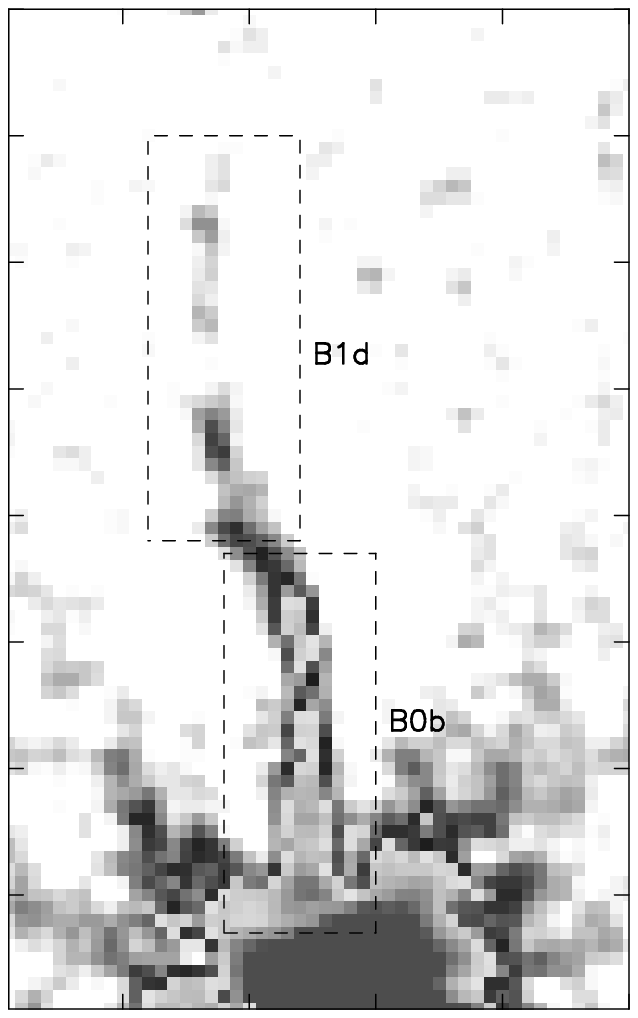}}\hfill
  \resizebox{0.5\textwidth}{!}{\plotone{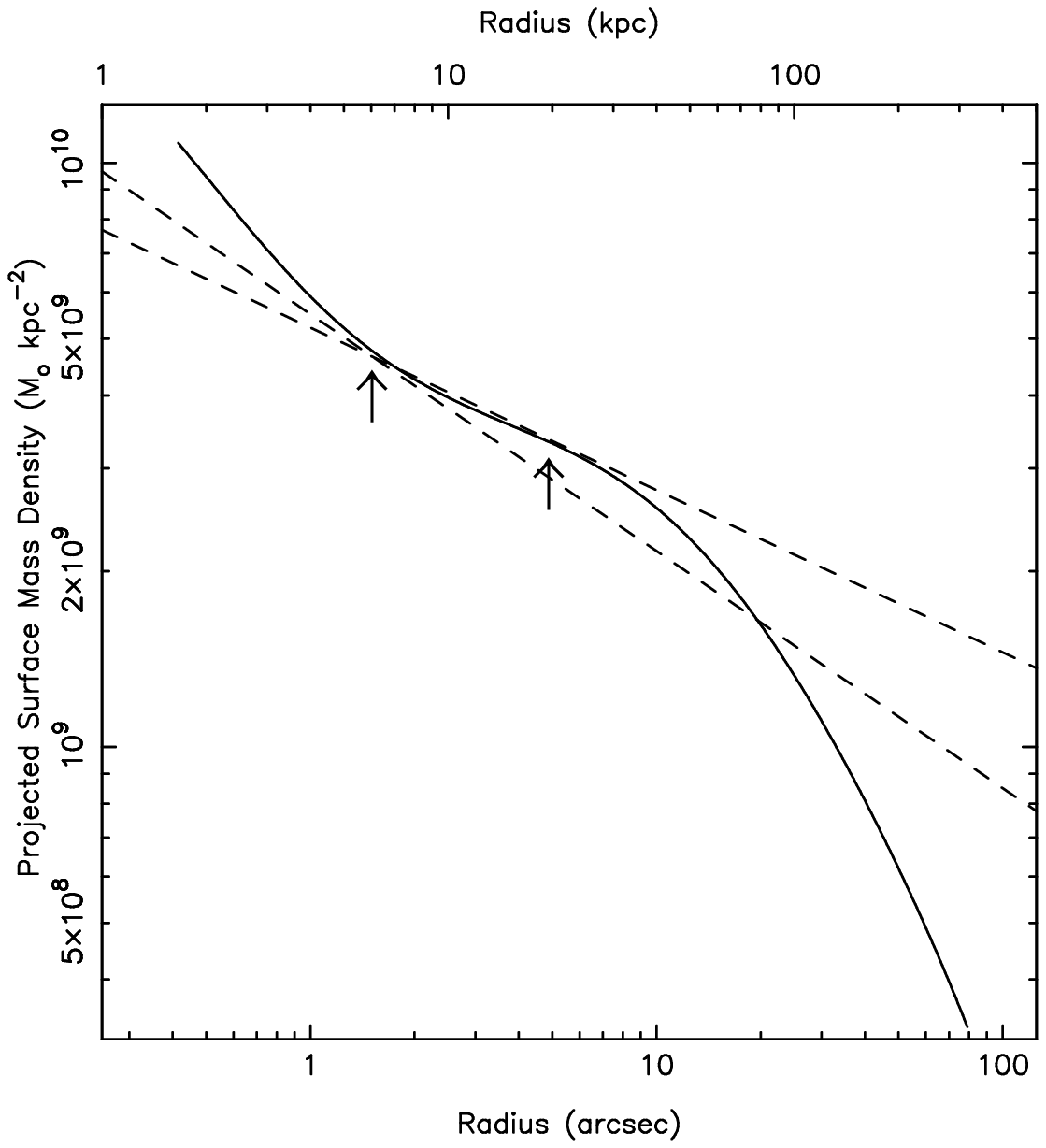}}
  \hspace*{0.05\textwidth}
  \caption{Left: Section of the WFPC2 image of Abell~383 showing the
  two radial arcs. Right: De-projected density profile of Abell~383,
  reconstructed from the strong lensing model. The positions of the
  radial arcs are marked by arrows. Both figures are taken from Smith
  et al.\ (2001).}
  \label{fig:a383-profile}
\end{figure}

\subsection{Weak Lensing}

Panoramic ground based images of all the clusters in the sample (with
the exception of Abell 773) were obtained during three observing runs
at CFHT during 1999/2000. The images were taken in the $B$, $R$ and
$I$ filters with the CFH12k camera, a mosaic camera of 12 CCD chips of
$2\mathrm{k}\times 4\mathrm{k}$ pixels each. The field of view of the
camera is $42\times28\,\mathrm{arcmin}^2$, or
$5.3\times3.5\,h^{-2}\,\mathrm{Mpc}^2$ at $z=0.2$. The data reduction
is described in Czoske (2002). The most critical step in the reduction
is the astrometric registration of the dithered exposures of a given
field and the assembly of the images from the 12 individual chips into
one contiguous image. Our registration pipeline achieves a final rms
deviation of object positions between the exposures of down to
$0\farcs01$, corresponding to 1/20 of the CFH12k pixel scale.

Galaxy shape measurements are done using the \texttt{im2shape}
software by S.\ Bridle which models the galaxy and PSF shapes as the
sum of 2-dimensional Gaussian profiles. Mass maps are constructed from
the shear field of background galaxies using \texttt{LensEnt}, a
method based on the maximum entropy principle (Bridle et al.\ 1998,
Marshall et al.\ 2002). Fig. 4 shows
the surface mass density field around Abell 68, reconstructed from
CFH12k $R$-band image. Apart from the cluster itself the
reconstruction reveals a number of additional mass concentrations,
most of which correspond to overdensities in the galaxy
distribution. Redshifts for these overdensities are needed in order to
calibrate the absolute masses associated with them and to establish
a possible connection to the main cluster. The numbers of
serendipitously found mass concentrations in our fields is consistent
with the dedicated survey by Miyazaki et al.\ (2002). 

\begin{figure}
  \centering
  \resizebox{!}{0.35\textheight}{\plotone{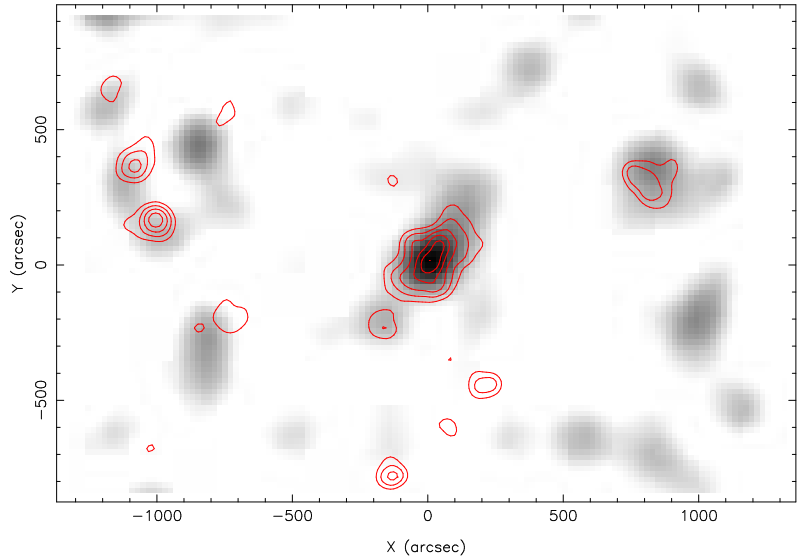}}
  \caption{The surface mass density of the field around Abell 68, as
    reconstructed from a weak lensing analysis of a deep $R$ band
    CFH12k image is shown in grey scale. The over-plotted contours give
    the galaxy number density in
    the field.}
  \label{fig:reconstruction}
\end{figure}

\subsection{Redshift distribution}
\label{ssec:redshifts}

Gravitational lensing yields direct and robust mass measurements, but
it is important to note that these are always weighted sums over all
mass contributions along the line of sight between the observer and
the sources, and therefore care must be taken if one wants to
interpret two-dimensional lensing masses in terms of three-dimensional
cluster masses. A striking example of how the projection of two
clusters along the line of sight can indicate a spurious large cluster
mass has been provided through the wide-field spectroscopic survey of
the cluster Cl0024+1654 at $z=0.395$ by Czoske et al.\ (2001, 2002).
In this case the complicated line-of-sight structure of the cluster
was only revealed through the redshift distribution and was not
visible in any other type of observation. In order to investigate the
presence of substructure in our cluster sample we obtain sizable
samples of galaxy redshifts for all of our clusters. The redshift
histograms are shown in Fig.\ 5. As can be seen,
several of our clusters do indeed show signs of substructure which has
to be taken into account in the interpretation of the gravitational
lensing results. In the cases where no substructure is present the
redshift distribution provides a valuable additional independent mass
estimate for the cluster.

\begin{figure}[htbp]
  \label{fig:histograms}
  \centering
  \resizebox{!}{0.35\textheight}{\plotone{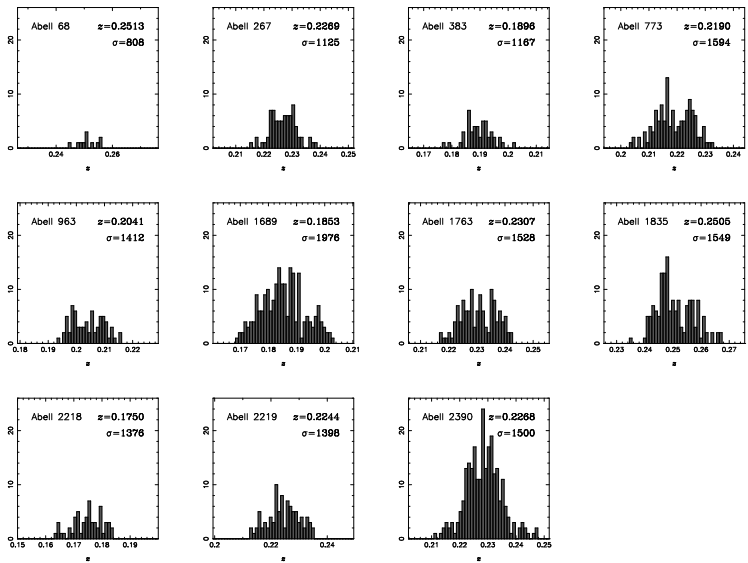}}
  \caption{Redshift histograms for 11 X-ray luminous clusters at
  $z\sim0.2$ (Czoske et al.\ 2003)}
\end{figure}

\section{Other XBACs/BCS follow-up projects}
\label{sec:follow-up}

\subsection{XBACS/BCS VLT/Gemini survey}
\label{ssec:cypriano}

In this project (Cypriano et al. 2002) shallow exposures of 24
clusters at $z>0.05$ and $L_{\rm X} >
5\times10^{44}\,\mathrm{erg\,s^{-1}}$ are observed in the $V$, $R$ and
$I$ bands with VLT/FORS1; a further 8 clusters were observed with
Gemini/GMOS in $g$, $r$ and $i$. The sample lends itself to test
predictions on the numbers of gravitational arcs expected depending on
the mass profile of the cluster. As shown by Meneghetti et al.\ 
(2002), profiles that are shallower than a singular isothermal sphere
(such as the NFW profile) lead to a marked decrease in the number of
arcs seen behind low-redshift clusters ($z<0.2$), whereas the
isothermal sphere would have a roughly constant cross section for arc
formation down to very low redshifts. First results from the XBACS
subsample show that the number of arcs per cluster is indeed smaller
for the low-redshift clusters ($z\sim0.1$) than for the
higher-redshift clusters ($z\ga0.2$) by a factor $\sim10$, indicating
that cluster profiles are shallower than isothermal (Fig.\ 6, Cypriano
et al.\ 2002).

\begin{figure}[htbp]
  \label{fig:cypriano}
  \centering
  \resizebox{0.45\textwidth}{!}{\plotone{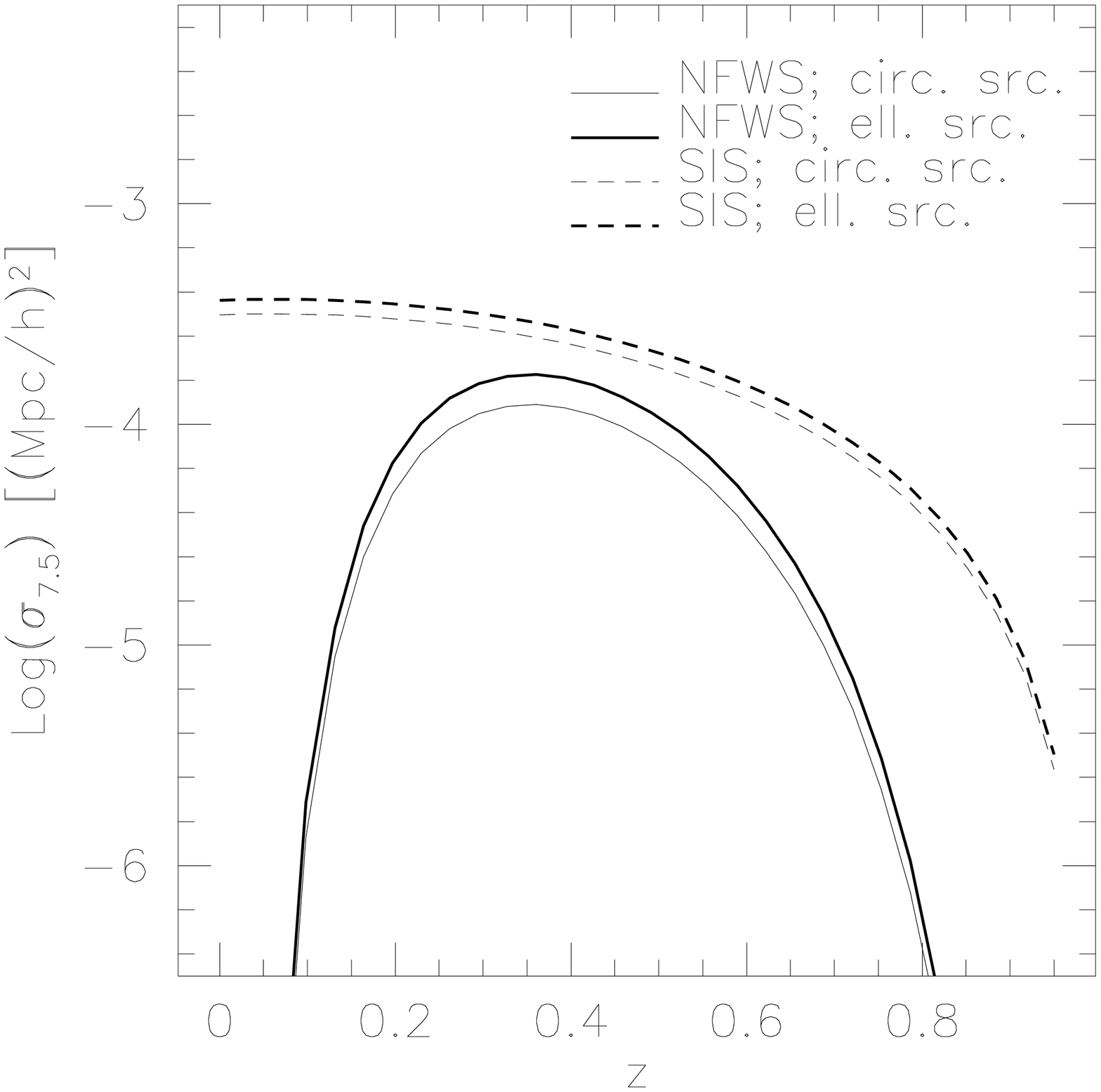}}\hfill
  \resizebox{0.45\textwidth}{!}{\plotone{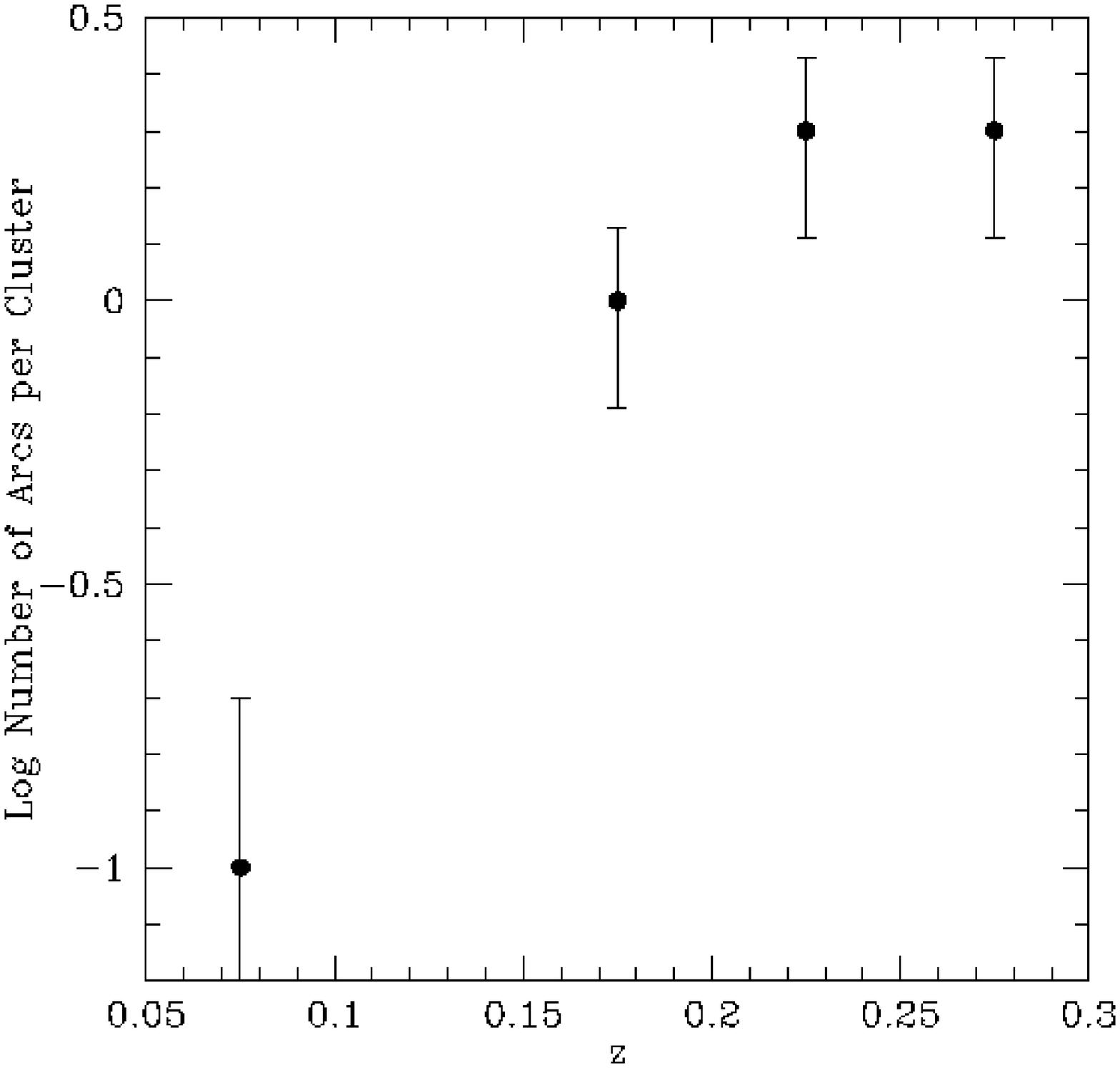}}
  \caption{Left: predictions for the cross-section for giant arc formation as a
    function of redshift for different halo mass models (from
    Meneghetti et al.\ 2002). Right: number of arcs per cluster found in the
    VLT survey for 24 XBACS clusters. The number of arcs decreases
    markedly for the low-redshift clusters, indicating that their mass
    profiles are shallower than isothermal.}
\end{figure}

\subsection{HST BCS Snapshot Survey}
\label{ssec:snapshot}

An HST Snapshot Survey of the central galaxies of more than 50
clusters taken from the BCS is being conducted by A. Edge and
collaborators. The primary goal of this project is to study in detail
the optical morphology on small scales of a complete sample of these
most massive stellar systems. However, these shallow images can also
used to detect gravitational arcs and arclets in the centres of these
clusters, thus providing an unbiased sample suitable for arc
statistics.

\subsection{Keck spectroscopy of central cluster galaxies}
\label{ssec:keck}

In a project led by T. Treu and R. Ellis, Keck spectroscopy of central cluster
galaxies and surrounding arcs and arclets is used to compare the
velocity dispersion of stars in the central galaxies with masses
derived from lensing. Arc redshifts are essential for accurate mass
determinations from strong lensing; since arcs have low surface
brightness 10m class telescopes are necessary in order to be able to
determine their redshifts. Comparing stellar velocity dispersions to
lensing masses allows an assessment of the importance of the mass
contribution from stars in the cluster centre relative to the dark
matter contribution (Sand, Treu \& Ellis 2002).

\section{The Ultimate MACS Survey}
\label{sec:MACS}

The MACS survey (Ebeling et al.\ 2001) is an all-sky X-ray survey
based on the \textsc{Rosat} All-Sky Survey data that for the first
time combines sufficient depth and sky-coverage to find the most X-ray
luminous clusters at high redshift ($z>0.3$). Previous surveys that
were looking for high-redshift clusters, such as EMSS or WARPS, were
restricted to small sky-area and thus missed the rare high-$L_{\rm X}$
objects, as shown in Fig.\ 7. At present, 119 MACS
clusters at $z>0.3$ are known; most of these are to be observed with
\textsc{Chandra}. MACS clusters are being imaged with
\textsc{Subaru}/Suprime to conduct weak lensing analyses; in addition,
spectroscopic follow-up observations are being conducted to
investigate cluster dynamics and to look for
substructure. Unfortunately, there is no HST imaging yet. 

\begin{figure}[htbp]
  \label{fig:macs}
  \centering
  \resizebox{!}{0.3\textheight}{\plotone{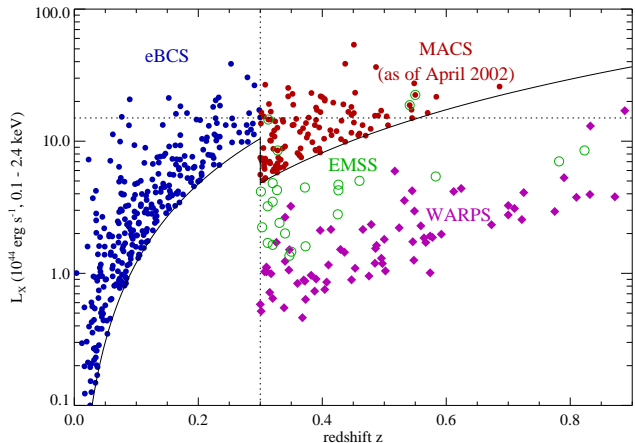}}
  \caption{$L_{\rm X}$--$z$ distribution of the extended BCS (Ebeling
  et al.\ 2000) at $z<0.3$ and of the preliminary MACS sample (119
  clusters) at $z>0.3$. By design, MACS finds the high-redshift
  counterparts of the most X-ray luminous clusters in the local
  universe. Also shown are the EMSS and WARPS samples.}
\end{figure}

\section{Conclusions}
\label{sec:conclusions}

Clusters of galaxies are complex systems. Each cluster is an
individual and we need to study large homogeneous samples in order to
properly assess the diversity of the cluster populations. We have
presented a number of on--going projects which aim at studying such
samples of the most massive clusters from a variety of viewpoints,
using all currently available observational techniques. These
parallel/complementary approaches, combining X-ray, lensing, dynamics,
SZ etc., are important in order to investigate the physics of
clusters. The aims of these studies are to understand the mass
distribution of clusters and to probe the cluster mass evolution with
redshift from $z=0$ to $z\la0.7$; to understand the importance of the
physical processes in clusters and the relation of observable
properties to mass; to assess the frequency of substructure and
mergers and how they affect the statistical properties of cluster
catalogues; and, finally, to relate these results to the cosmological
framework (Smith et al.\ 2003). We plan to conduct studies similar to
the $z=0.2$ sample (Sect.\ 2) at different redshifts, $z=0.1$, $0.4$
and $0.8$, the latter based on subsamples of MACS.

\acknowledgements This contribution presents work done in
collaboration with many colleagues: Ian Smail, Graham Smith, Harald
Ebeling, Sarah Bridle, Alastair Edge, Luis Campusano, L. Sodr\'e,
Eduardo Cypriano, Phil Marshall. OC thanks the organizers for giving
him the opportunity to attend this meeting.

\end{document}